\documentclass[usenatbib,onecolumn]{mn2e}
\def\gsim{\ \raise 3pt \hbox{$>$} \kern -8.5pt \raise -2pt \hbox{$\sim$}\ }
\def\lsim{\ \raise 3pt \hbox{$<$} \kern -8.5pt \raise -2pt \hbox{$\sim$}\ }

\newcommand{\pe}{_{\rm pe}}
\newcommand{\Be}{_{\rm Be}}
\newcommand{\st}{_{\rm st}}

\newcommand{\mmax}{_{\rm max}}

\usepackage{graphicx}

\begin{document}
   \title[DRL in jet shocks]{Diffusive  radiation in Langmuir turbulence
   produced by jet shocks}

\author[G. D. Fleishman and I. N. Toptygin]{G. D. Fleishman$^{1}$\thanks{E-mail:
gregory@sun.ioffe.ru} and I. N. Toptygin$^{2}$\thanks{E-mail:
cosmos@IT10242.spb.edu}\\
$^{1}$Ioffe Institute for Physics and Technology, St.Petersburg,
194021,  Russia; New Jersey
Institute of Technology, Newark, NJ 07102\\
$^{2}$State Polytechnical University, St.Petersburg, 195251, Russia}


\date{Accepted 2007 May 31. Received 2007 May 24; in original form 2007 March 19}

\pagerange{\pageref{firstpage}--\pageref{lastpage}} \pubyear{2007}

\maketitle

\label{firstpage}

\begin{abstract}
Anisotropic distributions of charged particles including two-stream
distributions give rise to generation of either stochastic electric
fields (in the form of Langmuir waves, Buneman instability) or
random quasi-static magnetic fields (Weibel and filamentation
instabilities) or both. These two-stream instabilities are known to
play a key role in collisionless shock formation, shock-shock
interactions, and shock-induced electromagnetic emission. This paper
applies the general non-perturbative stochastic theory of radiation
to study electromagnetic emission produced by relativistic
particles, which random walk in the stochastic electric fields of
the Langmuir waves. This analysis takes into account the cumulative
effect of uncorrelated Langmuir waves on the radiating particle
trajectory giving rise to angular diffusion of the particle, which
eventually modifies the corresponding radiation spectra. We
demonstrate that the radiative process considered is probably
relevant for emission produced in various kinds of astrophysical
jets, in particular,  prompt gamma-ray burst spectra, including
X-ray excesses and prompt optical flashes.
\end{abstract}


\section{Introduction}
Formation of collisionless astrophysical shocks, their interaction
with ambient medium and each other are tightly coupled with
two-stream instabilities giving rise to generation of either
electric or magnetic fields or both in the shock vicinity. In the
simplest kinetic version of the two-stream instability a tenuous
electron beam excites Langmuir waves resonantly (i.e., at
$k=\omega\pe/v_b$, where $k$ is the Langmuir wave vector,
$\omega\pe$ is the background plasma frequency, $v_b$ is the
velocity of the electron beam) on the linear stage of the
instability. However, even though these small-scale Langmuir waves
can initially be generated by the wave-particle resonance,
non-linear wave-wave interactions will then transform the wave
energy to larger scales \citep{Kaplan_Tsytovich_1973}. Modern
numerical models support that very small-scale random fields are
generated initially at the shock front and then they evolve to
larger and larger scales \citep{Jaroschek_etal_2004,
Jaroschek_etal_2005,Nishikawa_etal_2003,Nishikawa_etal_2005}.
Curiously, most of the work focuses on generation of stochastic
\emph{magnetic} fields due to Weibel or filamentation
instabilities, although highly efficient charged particle
acceleration occurring at the shock fronts necessarily requires
generation of stochastic \emph{electric} field  as well
\citep{Bykov_Uvarov_1993, Bykov_Uvarov_1999}. Indeed, numerical
simulations of relativistic shocks
\citep[e.g.,][]{Nishikawa_etal_2005} reveal very strong
fluctuations of the electric charge $\rho$, which are tightly
linked (${\bf kE} =4\pi \rho$) with the longitudinal electric
fluctuations.


Quasilinear approach performed mainly analytically
\citep{Bret_etal_2005} in 3-dimensional case shows that oblique
modes dominate both purely longitudinal  and purely transverse
modes. This is true also for the case when a magnetic field parallel
the beam velocity direction is included \citep{Bret_etal_2006} even
though this magnetic field can completely suppress the purely
transverse filamentation and Weibel modes.  Both magnetic and
electric fields are produced in the oblique modes and either
electric or magnetic energy density can dominate depending on
situation.  Being excited these waves are subject of sophisticated
evolution due to interaction with each other and with charged
particles \citep{Silva_2006}. These processes are deeply nonlinear
and highly sensitive to the input parameters of the particular
model.

Accordingly, there can be regimes when magnetic (electric)
inhomogeneities are dominant during the entire evolution of the
system and the effect of the electric (magnetic) inhomogeneities
can be discarded. For example, \cite{Dieckmann_2005} considered an
ultrarelativistic regime of the two-stream instability when a
broad spectrum of Langmuir waves (including non-resonant,
large-scale, ones $k \ll \omega\pe/v_b$) is excited and persist
for a long time in the plasma. However, different regimes when
initially dominant magnetic turbulence then gives a way to the
electric turbulence and vice versa are also possible \citep[for a
more instructive review, see][]{Silva_2006}. It would, therefore,
be highly desirable to have observational tests allowing to
distinguish between these different regimes, which can only come
from adequate interpretation of electromagnetic emission recorded
from an object.

Surprisingly, the radiation arising as relativistic charged
particles interact with stochastic electric fields of the Langmuir
turbulence has not been studied yet in sufficient detail, even
though a number of important and useful results have been obtained
for the past 40 years. Initially, this radiative process was
considered by \cite{Gailitis_Tsytovich_1964} \citep[see
also][]{Kaplan_Tsytovich_1973} for the case of small-scale
Langmuir waves ($k \gg \omega\pe/c$, $c$ is the speed of light)
and then the theory was extended to the case of the spatially
uniform ($k = 0$) electric fluctuations with the plasma frequency
\citep{Tsytovich_Chikhachev_1969}. The issue of the characteristic
frequency of this emission process was addressed by
\cite{Melrose_1971}, who call it after \cite{Colgate_1967}
"electrostatic bremsstrahlung" in contrast to "magneto
bremsstrahlung", while \cite{Schlickeiser_2003} considered the
total power of this process in a monochromatic approximation.

\cite{Getmantsev_Tokarev_1972} and \cite{Chiuderi_Veltri_1974}
demonstrated that an ensemble of relativistic electrons interacting
with the Langmuir turbulence would produce the emission with the
same spectral index as for standard synchrotron radiation. The
degree of polarization (for the case of highly anisotropic,
one-dimensional Langmuir turbulence) was found to be as large as for
synchrotron radiation in a uniform magnetic field
\citep{Tsytovich_Chikhachev_1969,Kaplan_Tsytovich_1973,Windsor_Kellogg_1974}.
Therefore, one might conclude that these two emission processes are
undistinguishable observationally
\citep[e.g.,][]{Windsor_Kellogg_1974}. We believe, this conclusion
is too straightforward. Indeed, for a narrow energy distribution of
emitting electrons or for a broad (power-law) distribution with
sufficiently sharp low- and/or high- energy cut-off, the specral
shapes of the emissions can be remarkably different, enabling us to
distinguish between various radiative processes.

However, full description of possible spectral regimes of this
radiative process has not been presented yet, although the general
theoretical foundation for this is readily available
\citep{Topt_Fl_1987,Topt_etal_1987,Fl_2005b,Fl_2005a}. In
particular, \cite{Topt_Fl_1987} clearly demonstrated that the
emission in the presence of random electric fields berries a lot of
general similarities to the emission in the random magnetic fields
\citep[so called diffusive synchrotron radiation,
DSR,][]{Fl_2005b,Fl_2005a}, although the exact expression for the
electron scattering rate should be adjusted accordingly to properly
take into account the polarization and dispersion of the Langmuir
waves. The physics lying behind these similarities relates to the
diffusive random walk of the emitting particle as it is being
randomly scattered by either electric or magnetic irregularities. To
emphasize that this diffusive motion is a key property to describe
the emission correctly, we will refer to this emission process as
"Diffusive Radiation in Langmuir turbulence" (or DRL) to be
distinguished from its cousin DSR. So far, there is a number of
terms suggested for this emission process, e.g., inverse Compton
scattering of the plasma waves, inverse plasmon scattering, and
electrostatic bremsstrahlung, although none of them is commonly
accepted, since none of the titles is indicative enough for the
radiative process considered.  It is worth noting that the term
"electrostatic bremsstrahlung" is not well suited for the emission
in the Langmuir waves because a more usual "bremsstrahlung" is even
more "electrostatic", than the DRL process discussed here.

%

\section{Perturbation theory of DRL}
\label{S_Pert_DRL}

Perturbative treatment of electromagnetic emission by a charged
particle assumes that the particle moves rectilinearly with constant
velocity but takes into account non-zero acceleration of the
particle in the external field. This perturbative treatment is
widely used because of its simplicity. Typically, one calculates
first the particle acceleration ${\bf w}(t)$ due to a given field
along the rectilinear trajectory and then uses this expression
obtained for ${\bf w}(t)$ to find the radiation spectrum. In the
case of a random external field, however, when ${\bf w}(t)$ is also
a random function of time $t$ it is more convenient to express the
radiation intensity via spatial and temporal spectrum of the
external electric and/or magnetic field.

Within theoretical formulation proposed by \cite{Fl_2005a} the
spectral and angular distribution of the emission produced by a
single particle with the Lorenz-factor $\gamma$ in a plasma with
random field has the form:
\begin{equation}
\label{cal_E_wFn_perp_2}
  W_{{\bf n},\omega}^{\bot}=\frac{(2\pi)^3Q^2}{M^2c^3\gamma^2V}
  \left(\frac{\omega}{\omega
  '}\right)^2
  \left[1-\frac{\omega}{\omega' \gamma_*^2} + \frac{\omega^2}{2\omega'^2 \gamma_*^4}
\right]
    \int  dq_0 d{\bf q}
  \delta(\omega'-q_0+{\bf qv}) \mid
 {\bf F}_{q_0, {\bf q}\bot} \mid^2.
\end{equation}
where $\gamma_* =
\left(\gamma^{-2}+\frac{\omega\pe^2}{\omega^2}\right)^{-1/2}$, $Q$,
$M$, and $\gamma \gg 1$ are the charge, mass, and Lorenz-factor of
the emitting particle, $V$ is the volume of the emission source,
${\bf F}_{q_0, {\bf q}\bot}$ is the Fourier component at the
frequency $q_0$ and the wave vector ${\bf q}$  of the Lorenz force
transverse to the emitting particle velocity,
\begin{equation}
\label{om_prime_def}
  \omega'=\frac{\omega}{2}\left(\gamma^{-2}+\theta^2 + \frac{\omega\pe^2}{\omega^2}
  \right),
\end{equation}
$\theta \ll 1$ is the angle between the wave vector of emitted wave
${\bf k}= k {\bf n}$ and the particle velocity vector ${\bf v}$,
$\omega$ is the emitted frequency. Contribution $W_{{\bf
n},\omega}^{\bot}$ (marked with the superscript $\bot$) is provided
by a component of the particle acceleration transverse to the
particle velocity. In case of the electric (in contrast to magnetic)
field, there is also a component of the acceleration along the
particle velocity. The corresponding contribution has the form
\begin{equation}
\label{cal_E_w_par_2}
  W_{{\bf n},\omega}^{\|}=\frac{2(2\pi)^3Q^4}{M^2c^3\gamma^6V}  \left(\frac{\omega}{\omega
  '}\right)^3
  \left[1-\frac{\omega}{2\omega' \gamma_*^2}
\right]
    \int  dq_0 d{\bf q}
  \delta(\omega'-q_0+{\bf qv}) \mid
 {\bf E}_{q_0, {\bf q}\|} \mid^2,
\end{equation}
which is typically small by a factor $\gamma^{-2}$ compared with the
transverse contribution. Nevertheless, there exist special cases
(e.g., a particle moving along one-dimensional turbulent electric
field) when the transverse contribution is zero or very small and
the parallel contribution comes to play. For example, this is the
case when a charge particle moves along the electric field of
one-dimensional Langmuir turbulence \citep{Fl_Topt_2007}.

However, below we will consider only the transverse contribution,
which is indeed the dominant one in most of the cases.
Accordingly, the spectral distribution of the radiated energy is
given by integration of (\ref{cal_E_wFn_perp_2}) over the full
solid angle $d\Omega=\sin\theta d\theta d\varphi \approx 2\pi
d(\omega'/\omega)$:
\begin{equation}
\label{cal_E_w_F_perp_3}
  W_{\omega}^{\bot}=\frac{(2\pi)^4Q^2}{M^2c^3\gamma^2V}
  \int_{1/2\gamma_*^2}^{\infty} d\left(
  \frac{\omega'}{\omega} \right)
  \left(\frac{\omega}{\omega '}\right)^2
  \left[1-\frac{\omega}{\omega'\gamma_*^2} +
  \frac{\omega^2}{2\omega'^2\gamma_*^4  }
\right]
    \int  dq_0 d{\bf q}
  \delta(\omega'-q_0+{\bf qv}) \mid
 {\bf F}_{q_0, {\bf q}\bot} \mid^2
\end{equation}
similar to the DSR case \citep{Fl_2005a}, but with the Lorenz
force ${\bf F} = Q {\bf E}$ specified by electric ${\bf E}$ in
place of magnetic field.

\subsection{One-wave approximation in the DRL theory} \label{S_One_wave_DRL}

Let's consider first the simplest case when there is only one long
Langmuir wave with $k_0 c \ll \omega\pe$, where ${\bf k}_0$ is the
wave-vector and $\omega\pe$ is the frequency of the Langmuir wave,
which was described long ago by \cite{Tsytovich_Chikhachev_1969}.
Here, the spatial and temporal spectrum of the electric field in
this Langmuir wave takes a simple form
\begin{equation}
\label{E_tran_Lang_2}
     \mid {\bf E}_{\bot}(q_0,{\bf q})
   \mid^2
   =A_E \frac{TV}{(2\pi)^4}
 \delta({\bf q}-{\bf k}_0)( \delta(q_0-\omega\pe)+\delta(q_0+\omega\pe)),
\end{equation}
where $A_E$ is specified by the energy density of the electric
field in the wave. For example, for spatially uniform electric
oscillations
\begin{equation}
\label{E_Lang_hom}
 {\bf E}={\bf E}_0 \cos \omega\pe t,
\end{equation}
we have $A_E=\mid {\bf E}_{0\bot} \mid^2/4$ .

Calculation of the radiation intensity is extremely easy in case of
field spectrum (\ref{E_tran_Lang_2}). Indeed, substituting
(\ref{E_tran_Lang_2}) into (\ref{cal_E_wFn_perp_2}) and taking the
integrals over frequency $q_0$ and wave-vector ${\bf q}$ using the
$\delta$-functions, we obtain the emission intensity (i.e., the
energy emitted per unit time per unit frequency interval per unit
solid angle) by dividing (\ref{cal_E_wFn_perp_2}) over $T$:
\begin{equation}
\label{I_Lang_nw_2}
  I_{{\bf n},\omega}^{\bot}=\frac{A_EQ^4}{2\pi M^2c^3\gamma^2}  \left(\frac{\omega}{\omega
  '}\right)^2
  \left[1-\frac{\omega}{\omega'}\left(\gamma^{-2}+\frac{\omega\pe^2}{\omega^2}\right) +
  \frac{\omega^2}{2\omega'^2
  }\left(\gamma^{-2}+\frac{\omega\pe^2}{\omega^2}\right)^2
\right] 
   \delta(\omega'-\omega\pe+{\bf k}_0 {\bf v})
   .
\end{equation}

Let's discuss this emission intensity in more detail. Recall
(\ref{om_prime_def}) that $\omega'$ is a function of the emission
angle $\theta$. On the other hand, because of the
$\delta$-function in (\ref{I_Lang_nw_2}), $\omega'$ is a fixed
number for a given set of values $\omega\pe$, ${\bf k}_0$, and
${\bf v}$:
\begin{equation}
\label{Lang_Rad_cond}
 \omega' = \omega\pe-{\bf k}_0 {\bf v}.
\end{equation}
Therefore, there is strict correlation between the emission
frequency and direction, thus, only one distinct frequency can be
emitted along a given direction.

Interestingly,  emission intensity (\ref{I_Lang_nw_2}) depends only
weakly on the actual ${\bf k}_0$ value as long as the condition $k_0
c \ll \omega\pe$ holds. In particular, for a long Langmuir wave the
radiation intensity is almost the same as for the spatially uniform
temporal oscillations of the electric field with the plasma
frequency. The radiation intensity into full solid angle (which is
the differential spectral power, i.e., the energy emitted per unit
time per unit frequency range) is given by simple integration of
(\ref{I_Lang_nw_2}) over $d(\omega'/\omega)$ with the use of the
$\delta$-function. It does depend on ${\bf k}_0$ only weakly and for
${\bf k}_0=0$ has the form
\begin{equation}
\label{I_Lang_w_2}
  I_{\omega}^{\bot}=\frac{Q^4{\bf E}^2_{0\bot}}{4 M^2c^3\gamma^2}  \frac{\omega}{\omega\pe^2}
  \left[1-\frac{\omega}{\omega\pe}\left(\gamma^{-2}+\frac{\omega\pe^2}{\omega^2}\right) +
  \frac{\omega^2}{2\omega\pe^2
  }\left(\gamma^{-2}+\frac{\omega\pe^2}{\omega^2}\right)^2
\right], \qquad \omega < 2\omega\pe\gamma^2
   .
\end{equation}
in full agreement with the result of
\cite{Tsytovich_Chikhachev_1969}.

The dispersion of the plasma (terms $\omega\pe^2/\omega^2$) has
almost no effect on the radiation intensity. The spectrum has
unique asymptote $I_\omega \propto \omega^1$ at low frequencies,
$\omega \ll \omega\pe\gamma^2$ with a peak at $\omega_{max} =
2\omega\pe\gamma^2$. The emission intensity vanishes at higher
frequencies.

Given that the radiation intensity for the case of single long
Langmuir wave has only a weak dependence on the actual wavelength,
it is tempting to extrapolate the presented results (in
particular, spectrum (\ref{I_Lang_w_2}) to the case when an
ensemble of many long Langmuir waves exists in the source volume.
Below we will see, however, that it is fundamentally incorrect,
because the presence of the broad spatial spectrum of electric
fluctuations makes a great difference compared with the one-wave
case; general trends here are similar to those for the DSR in
random magnetic fields \citep{Fl_2005a}.

\subsection{Perturbation theory for broad spatial spectrum} \label{S_Broad_pert_DRL}

Following the derivation given by \cite{Fl_2005a}, it is easy to
find
\begin{equation}
\label{E_corr}
       \mid {\bf E}_{q_0,{\bf q}\bot} \mid^2
   = \frac{TV}{(2\pi)^4}\left(\delta_{\alpha\beta} -
\frac{v_{\alpha}v_{\beta}}{v^2}\right) K_{\alpha \beta}(q_0,{\bf
q})
\end{equation}
where $K_{\alpha\beta}(q_0,{\bf q}) = C_{\alpha\beta} K(q_0,{\bf
q})$; $C_{\alpha\beta}= q_{\alpha}q_{\beta}/q^2$ for the Langmuir
turbulence since the electric field vector is directed along the
wave vector in the Langmuir waves,  while $K(q_0,{\bf q})$ is the
temporal and spatial spectrum of the Langmuir turbulence.
Substituting (\ref{E_corr}) into general expression
(\ref{cal_E_wFn_perp_2}) and dividing the result by $T$ we obtain
spectral and angular intensity of DRL per unit time, which is
convenient to write down in the form:

\begin{equation}
\label{I_perp_q_gen_n}
  I_{{\bf n},\omega}=\frac{Q^2}{2\pi^2 c}
    \left(\frac{\omega}{\omega '}\right)^2
  \left[1-\frac{\omega}{\omega'\gamma_*^2} +
  \frac{\omega^2}{2\omega'^2\gamma_*^4  }
\right]
    q_L(\omega,\theta),
\end{equation}
where $q_L(\omega,\theta) \equiv q_L(\omega')$ is the effective
scattering rate of the relativistic particle by the Langmuir
turbulence, which plays a key role within the full non-perturbative
treatment of the DRL:
\begin{equation}
\label{q_delta_main}
 q_L(\omega,\theta) =  \frac{\pi Q^2}{M^2c^2\gamma^2}
  \int dq_0 d{\bf q}\left(\delta_{\alpha\beta} -\frac{v_\alpha
v_\beta}{v^2}\right) K_{\alpha \beta}(q_0,{\bf q})
\delta(\omega'-q_0 +{\bf q}{\bf v})
 .
\end{equation}

To perform further calculations we have to specify the form of
$K_{\alpha\beta}(q_0,{\bf q})$. Initially, the streaming instability
can give rise to highly anisotropic turbulent spectrum with a
limiting case of purely one-dimensional turbulence, although
randomization of the wave vectors will later result in more
isotropic turbulence patterns. DRL generated in the presence of
one-dimensional Langmuir turbulence is specifically discussed by
\cite{Fl_Topt_2007}. By contrast, here we assume that the Langmuir
wave vectors are isotropically distributed and the spectrum
$K(q_0,{\bf q})$ can be approximated by a power-law above certain
critical value $k_0$:
\begin{equation}
\label{E_tran_Lang_spectr}
     K(q_0,{\bf q})
   =  \frac{a_\nu k_0^{\nu-1} \left<E_L^2\right> q^2}{(k_0^2+q^2)^{\nu/2+2}}
  \delta(q_0-\omega\pe).
\end{equation}
Here, the presence of the $\delta$-function is related to the
assumption that the electric turbulence is composed of Langmuir
waves all of which oscillate in time with the same frequency
$\omega\pe$; the normalization constant $a_\nu$ is set up by the
condition $\int K(q_0,{\bf q}) dq_0 d{\bf q} =
\left<E_L^2\right>$, where  $\left<E_L^2\right>$ is the mean
square of the electric field in the Langmuir turbulence.

Now, substituting (\ref{E_tran_Lang_spectr}) into
(\ref{q_delta_main}), taking the integrals over $dq_0$ and
$d\cos\theta_q$ with the use of two available $\delta$-functions in
(\ref{q_delta_main}), and then integrating over $dq$, we find
\begin{equation}
\label{q_theta_integrated}
 q_L(\omega,\theta) =  \frac{4\pi^2 a_\nu k_0^{\nu-1} \omega\st^2}{\nu(\nu+2) c \gamma^2}
  \left\{\left(\frac{\omega' -\omega\pe}{c}\right)^2
   +k_0^2\right\}^{-\nu/2}
 ,
\end{equation}
where $\omega_{st}=Q\left< E_{L}^2\right>^{1/2}/Mc$. Substitution of
(\ref{q_theta_integrated}) into (\ref{I_perp_q_gen_n}) yields:

\begin{equation}
\label{I_Lang_w_spectr}
  I_{{\bf n},\omega}^{\bot}=\frac{2a_\nu k_0^{\nu-1} \omega\st^2  Q^2 }
  {\nu(\nu+2) c^2\gamma^2}  \left(\frac{\omega}{\omega
  '}\right)^2
  \left[1-\frac{\omega}{\omega'}\left(\gamma^{-2}+\frac{\omega\pe^2}{\omega^2}\right) +
  \frac{\omega^2}{2\omega'^2
  }\left(\gamma^{-2}+\frac{\omega\pe^2}{\omega^2}\right)^2
\right] 
   \left\{\left(\frac{\omega' -\omega\pe}{c}\right)^2
   +k_0^2\right\}^{-\nu/2}
   .
\end{equation}
Apparently, spectrum (\ref{I_Lang_w_spectr}) looks rather
differently from that in case of a single Langmuir wave
(\ref{I_Lang_w_2}). In particular, no $\delta$-function enters
(\ref{I_Lang_w_spectr}), thus, a continuum spectrum rather than
distinct frequencies is emitted along any direction. Clearly,
there remains a distinct contribution to the emission intensity
when $\omega' \approx \omega\pe$. However, the range of the
parameter space where this resonant condition holds is relatively
narrow, so the "non-resonant" contribution from the remaining part
of the parameter space where $\omega' \neq \omega\pe$ can easily
dominate the resonant contribution. To see this explicitly,
consider the radiation intensity into the full solid angle by
integration of (\ref{I_Lang_w_spectr}) over $d\Omega$ that yields

\begin{equation}
\label{I_case2_Lang_high}
 I_{\omega}^{\bot}  = 2\pi \int_{1/2\gamma_*^2}
  ^{\infty} I_{{\bf n}, \omega}^{\bot} d\left(\frac{\omega'}{\omega}\right)
 =\frac{8Q^2 \gamma_*^2}{3 \pi  c}  q_L(\omega),
\end{equation}
where
\begin{equation}
\label{q_plasm_gen_def}
 q_L(\omega)=
 \frac{3\pi^2 a_\nu k_0^{\nu-1} \omega\st^2  }
  {2\nu(\nu+2) c^3 \gamma^2 \gamma_*^2} \int_{1/2\gamma_*^2}
  ^{\infty} d\left(\frac{\omega'}{\omega}\right)
  \left(\frac{\omega}{\omega
  '}\right)^2
  \left[1-\frac{\omega}{\omega'}\left(\gamma^{-2}+\frac{\omega\pe^2}{\omega^2}\right) +
  \frac{\omega^2}{2\omega'^2
  }\left(\gamma^{-2}+\frac{\omega\pe^2}{\omega^2}\right)^2
\right] $$$$ 
   \left\{\left(\frac{\omega' -\omega\pe}{c}\right)^2
   +k_0^2\right\}^{-\nu/2}.
 \end{equation}
Here we express the radiation spectrum via the effective scattering
rate $q_L(\omega)=\overline{q_L(\omega,\theta)}$ averaged over the
emission angle $\theta$, which defines  the general non-perturbative
expressions of the radiation intensity. Note that this averaging is
performed with an appropriate weight described by the factor
$\left(\frac{\omega}{\omega '}\right)^2
  \left[1-\frac{\omega}{\omega'\gamma_*^2} +
  \frac{\omega^2}{2\omega'^2\gamma_*^4  }
\right]$, which enters Eqns (\ref{I_perp_q_gen_n}) and
(\ref{I_Lang_w_spectr}). This factor is important since $\omega'$
is a function of the emission angle $\theta$ according to equation
(\ref{om_prime_def}). Combination of (\ref{I_case2_Lang_high}) and
(\ref{q_plasm_gen_def}) yields finally:

\begin{equation}
\label{I_Lang_w_spectr_2}
  I_{\omega}^{\bot}=\frac{4\pi a_\nu k_0^{\nu-1} \omega\st^2  Q^2  }
  {\nu(\nu+2) c^2 \gamma^2} \int_{1/2\gamma_*^2}
  ^{\infty} d\left(\frac{\omega'}{\omega}\right)
  \left(\frac{\omega}{\omega
  '}\right)^2
  \left[1-\frac{\omega}{\omega'}\left(\gamma^{-2}+\frac{\omega\pe^2}{\omega^2}\right) +
  \frac{\omega^2}{2\omega'^2
  }\left(\gamma^{-2}+\frac{\omega\pe^2}{\omega^2}\right)^2
\right] $$$$ 
   \left\{\left(\frac{\omega' -\omega\pe}{c}\right)^2
   +k_0^2\right\}^{-\nu/2}
   .
\end{equation}

\begin{figure}
\hspace{-0.4in}
\includegraphics[bb=18 5 282 217,height=2.5in]{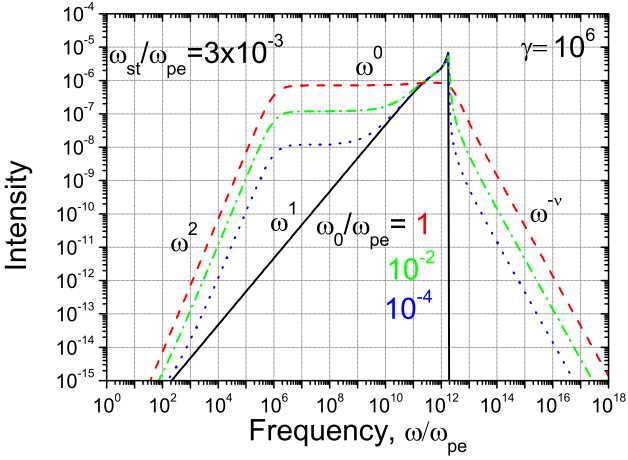} 
\caption{\small \it  Perturbative  DRL spectra produced by a
particle with $\gamma=10^6$ in a plasma with relatively weak
($\omega\st/\omega\pe=3\cdot10^{-3}$) long- and short- wave Langmuir
turbulence for various $\omega_0/\omega\pe$ ratios characterizing
the turbulence: $\omega_0/\omega\pe = 1$, dashed/red curve;
$\omega_0/\omega\pe = 10^{-2}$, dash-dotted/green curve;
$\omega_0/\omega\pe = 10^{-4}$, dotted/blue curve. The spectrum of
radiation in the presence of spatially uniform Langmuir oscillations
\citep{Tsytovich_Chikhachev_1969} is shown by solid/black curve for
comparison. The spectra contain a distinct spectral peak at $\omega
\sim 2\omega\pe \gamma^2$ for long-wave turbulence, which is absent
for the short-wave turbulence.
}
\label{Langmuir_Pert}
\end{figure}

This integral cannot be taken analytically in a general case, but
it is easy to plot corresponding spectra numerically. Figure
\ref{Langmuir_Pert} displays these spectra for a representative
set of involved parameters. There are prominent differences in the
DRL spectra in case of a broad spectrum of the Langmuir waves
compared with the one-wave spectrum ($\propto \omega^1$), which is
plotted in the figure for comparison. Even though we cannot
perform full analytical treatment of the  spectrum, we can
estimate the spectral shape in various frequency ranges. At low
frequencies $\omega \ll \omega\pe\gamma^2$, we can discard
$\omega'$ in (\ref{I_Lang_w_spectr_2}) everywhere in the braces
except narrow region of parameters when $\omega' \approx
\omega\pe$. This means that for $\omega \ll \omega\pe\gamma^2$ the
integral is composed of two contributions. The first of them, a
non-resonant one, arises from integration over the region, where
$\omega' \ll \omega\pe$. Here, the emission is beamed within the
characteristic emission angle of $\vartheta \sim \gamma^{-1}$
along the particle velocity. The integral converges rapidly, and
so it may be taken along the infinite region, which produces a
flat radiation spectrum, $I_\omega \propto \omega^0$, or $I_\omega
\propto \omega^2$ at lower frequencies, $\omega <
\omega\pe\gamma$. However, as far as $\omega'$ approaches
$\omega\pe$, a resonant contribution comes into play. Now, in a
narrow vicinity of $\omega\pe$, we can adopt
\begin{equation}
\label{del_Lang_approx}
     \left\{\left(\frac{\omega' -\omega\pe}{c}\right)^2
   +k_0^2\right\}^{-\nu/2} \propto \delta(\omega' -\omega\pe)
   ,
\end{equation}
which results in a single-wave-like contribution, $I_\omega
\propto \omega^1$. The full spectrum at $\omega <
\omega\pe\gamma^2$, therefore, is just a sum of these two
contributions.

At high frequencies, $\omega \gg \omega\pe\gamma^2$, the term
$\omega'$ dominates in the braces, so other terms can be
discarded. Thus, a power-law tail, $I_\omega \propto
\omega^{-\nu}$, typical for DSR high-frequency asymptote, arises
in this spectral range, where there was no emission et all in case
of a single Langmuir wave, \S \ref{S_One_wave_DRL}.

Let's compare the DRL spectra with the DSR spectra in case of
stochastic magnetic fields \citep{Fl_2005a}. If the Langmuir
turbulence consists of relatively small-scale waves, $\omega_0
\equiv k_0 c \gsim \omega\pe$ (see dashed curve in Figure
\ref{Langmuir_Pert}) then the shape of the spectrum is similar to
the DSR spectrum. However, there is a remarkable difference in the
case of the long-wave turbulence, $\omega_0 \ll \omega\pe$. Here,
a distinct spectral peak at $\omega = 2\omega\pe\gamma^2$ is
formed with the linear decrease of the spectrum with frequency,
therefore, the immediate vicinity of the spectral peak can indeed
be described within the single-wave  approximation as suggested by
\cite{Tsytovich_Chikhachev_1969}. At lower frequencies, however,
this falling part of the spectrum gives way to a flat spectrum,
which is entirely missing within the one-wave approach. Position
of the corresponding turning point depends on the
$\omega_0/\omega\pe$ ratio. It is worth emphasizing that the
deviations of the DRL spectrum from the single-wave spectrum is
prominent even for extremely long-wave turbulence, e.g., with
$\omega_0/\omega\pe=10^{-4}$ as in Figure \ref{Langmuir_Pert}.
Therefore, the presence of a broad turbulence spectrum results in
important qualitative change of the emission  mechanism, which
cannot generally be reduced to a simplified treatment relying on
the single-wave approximation with some rms value of the Langmuir
electric field.

\section{Non-perturbative treatment of DRL}
\label{S_Non_Pert_DRL}

The perturbative treatment of the emission produced by a
relativistic particle moving in the presence of random fields
breaks down sooner or later as soon as intensity of the field
and/or particle energy increase \citep[e.g.,][]{Fl_2005a}. To find
the applicability region of the perturbation theory applied above,
we should estimate the characteristic deflection angle of the
emitting electron on the emission coherence length
$l_c=2c\gamma_*^{2}/\omega$, where the elementary emission pattern
is formed. Similarly to \cite{Fl_2005a} consider a simple source
model consisting of uncorrelated cells with the size $l_0=2\pi
c/\omega_0$, each of which contains coherent Langmuir oscillations
with the plasma frequency $\omega\pe$. If $\omega_0 > \omega\pe$
then inside each cell the electron velocity will change by the
angle $\theta_0 \sim \omega\st/(\omega_0 \gamma)$, therefore, the
consideration given in \cite{Fl_2005a} applies. However, if
$\omega_0 \ll \omega\pe$, then the electric field vector will
change the direction approximately $(\omega\pe/\omega_0)$ times
during the time required for the particle to path through one
cell, thus, the net deflection angle will be reduced by this
factor $(\omega\pe/\omega_0)$ due to temporal oscillations of the
electric field in the Langmuir waves, therefore $\theta_0 \sim
\omega\st/(\omega\pe \gamma)$. Then, after traversing $N=l_c/l_0$
cells, the mean square of the deflection angle is $\theta_c^2
=\theta_0^2 N \sim \omega\st^2 \omega_0/(\omega\omega\pe^2)$. The
perturbation theory is only applicable if this diffusive
deflection angle is smaller than the relativistic beaming angle,
$\gamma^{-1}$, i.e., it is always valid at sufficiently high
frequencies $\omega
> \omega_* \equiv \omega\st^2 \omega_0 \gamma^2/\omega\pe^2$.
Note, that the bounding frequency $\omega_*$ increases with
$\omega_0$, while DSR displays the opposite trend.  The
perturbation theory will be applicable to the entire DRL spectrum
if the condition $\theta_c^2 < \gamma^{-2}$ holds for the
frequency $\omega\pe\gamma$ \citep{Fl_2005a}, where the coherence
length of the emission has a maximum. This happens for the
particles whose Lorenz-factors obey the inequality
\begin{equation}
 \label{DRL_criter}
\gamma \ll \omega\pe^3/(\omega\st^2\omega_0).
\end{equation}

We see, therefore, that generally, especially for relatively strong
Langmuir turbulence, the perturbative treatment is insufficient to
fully describe the radiation spectrum, thus, the non-perturbative
version \citep{Topt_Fl_1987,Fl_2005b} of the theory should be
explored. As demonstrated in \cite{Topt_Fl_1987} the same general
expressions for the radiation intensity produced in the presence of
stochastic electric fields are valid like in the random magnetic
fields, although the electron scattering rate by Langmuir waves
$q_L(\omega)=\overline{q_L(\omega,\theta)}$, which has already been
introduced in the previous section, will differ from that in case of
magnetic turbulence. As we will see below, all the differences
between DRL spectrum in Langmuir turbulence and the DSR spectrum in
magnetic turbulence are ultimately related to the difference in the
expressions for the scattering rate $q_L(\omega)$ for these two
cases.

Let's consider first the regime when there is no regular magnetic
field and, thus, the radiation spectrum is defined as
\citep{Fl_Biet_2007}
\begin{equation}
\label{I_DSR_stoch}
 I_{\omega}  = \frac{8Q^2 q_L(\omega)}{3 \pi  c} \gamma^2 \left(1+\frac{\omega\pe^2
 \gamma^2}{\omega^2}\right)^{-1} \Phi(s),
\end{equation}
where $\Phi(s)$ is the Migdal function \citep{Migdal,Migdal_1956}
\begin{equation}
\label{Fi_migdal_def*}
 \Phi(s)  = 24s^2 \int_0^{\infty}dt
 \exp(-2st)\sin (2st)
  \times 
 \left[\coth t -\frac{1}{t}\right], \qquad \hbox{with} \
  s  =  \frac{1}{8\gamma^2}
  \left(\frac{\omega}{q_L(\omega)}\right)^{1/2}\left(1+\frac{\omega\pe^2
 \gamma^2}{\omega^2}\right)
 \end{equation}
having rather simple asymptotes for large or small $s$ values:
\begin{equation}
\label{migdal_asympt}
 \Phi(s) \simeq 1, \hbox{if} \ s \gg 1,
 \qquad  \Phi(s) \simeq 6s, \hbox{if} \ s \ll 1.
\end{equation}

\begin{figure} 
\includegraphics[bb=18 5 282 217,height=2.3in]{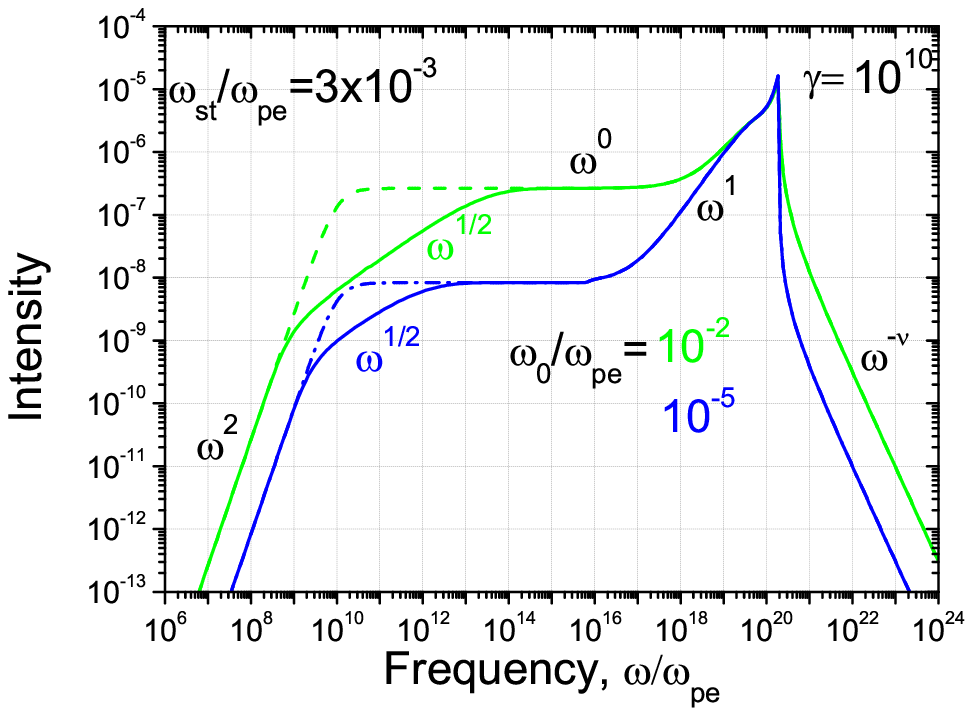} 
\qquad
\includegraphics[bb=18 5 282 217,height=2.3in]{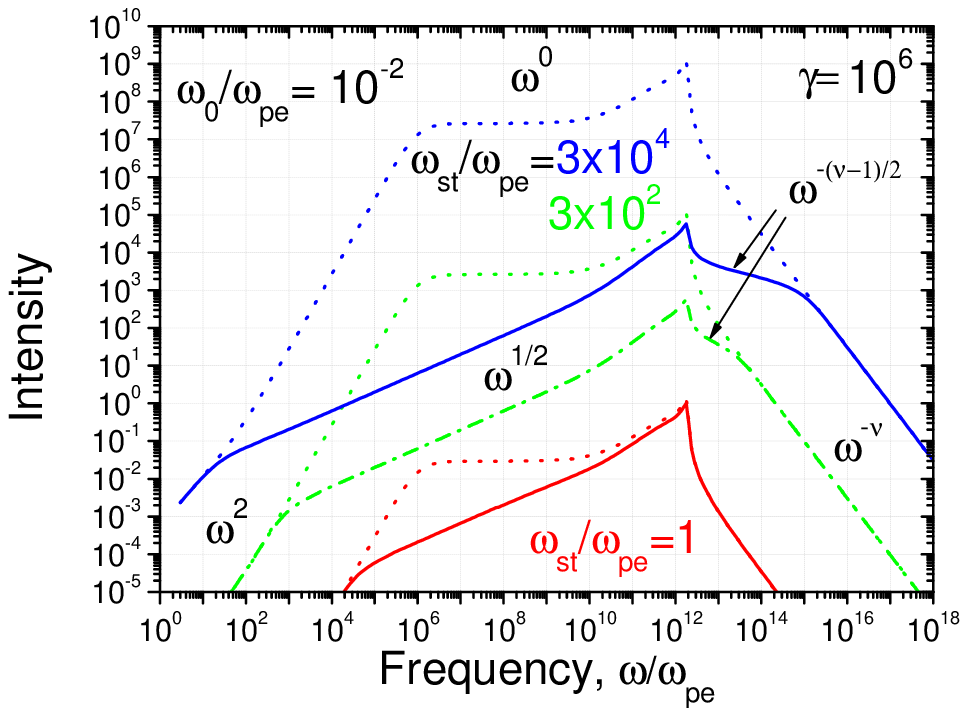}
\caption{\small \it DRL spectra produced by highly relativistic
particle with $\gamma = 10^{10}$ moving in a plasma with long-wave
Langmuir turbulence. Full non-perturbative spectra are shown by
solid curves, while the corresponding perturbative spectra are by
dashed and dash-dotted curves.}
%
\label{langmuir_weak} \caption{\small \it DRL in the presence of
strong long-wave Langmuir turbulence. Dotted curves are calculated
within the perturbation theory. Extremely strong suppression of the
DRL spectra compared with the perturbative ones is evident.
} \label{langmuir_strong}
\end{figure}

Figure \ref{langmuir_weak} presents the DRL spectra for the case of
relatively weak electric field, $\omega\st \ll \omega\pe$; the
dash-dotted curves are the corresponding perturbative spectra. The
non-perturbative effect (multiple scattering of the radiating
electron by Langmuir waves) modifies the spectrum around the
frequency $\omega\pe \gamma$ giving rise to asymptote  $I_\omega
\propto \omega^{1/2}$ in this spectral region. Note that the
frequency $\omega_*$, where the break from the $\propto \omega^0$ to
$\propto \omega^{1/2}$ asymptote occurs, increases with $\omega_0$
increase, while the DSR in the static random magnetic fields
\citep{Fl_2005a} displays the opposite trend.

However, as shown in \cite{Silva_2006} the electrostatic field in
the Langmuir waves generated at the shock front can be rather
strong, e.g., of the order of nonrelativistic wave breaking limit,
$\omega\st \approx \omega\pe$. In this case, the non-perturbative
treatment is important at the full frequency range below the
spectral peak at $2\omega\pe \gamma^2$, Figure
\ref{langmuir_strong}. For completeness of the possible DRL
regimes considered, Figure \ref{langmuir_strong} presents also the
DRL spectra for the (less realistic) case of a very strong random
electric field, $\omega\st \gg \omega\pe$. Here, the
non-perturbative spectrum deviates from the perturbative one even
at the frequencies above $2\omega\pe \gamma^2$, giving rise to a
suppressed spectrum $I_\omega \propto \omega^{-(\nu-1)/2}$
(compared with  the perturbative one $I_\omega \propto
\omega^{-\nu}$). At the lower frequencies, a very broad
non-perturbative power-law region of the spectrum, $I_\omega
\propto \omega^{1/2}$, is formed.

\section{DRL vs synchrotron radiation}
\label{S_DRL_vs_SR}

If both Langmuir turbulence and regular magnetic field are present
in the source, then both DRL and synchrotron emission are generated.
One could assume that the DRL is only relevant in extreme conditions
when the Langmuir turbulence energy density exceeds that of the
regular magnetic field \citep[e.g.,][]{Melrose_1971}. It is not the
case, however, because these two emission mechanisms are efficient
in differing frequency domains.

Consider joint effect of the Langmuir turbulence and a regular
magnetic field on the radiation spectrum. In this case full
equations    (30) from  \citep{Fl_2005b} together with the exact
expression for $q_L(\omega)$ given by Eq. (\ref{q_plasm_gen_def})
should be used. It is worth emphasizing that the condition
$\omega\Be \ll \omega\pe$, where $\omega\Be$ is cyclotron frequency
in the regular magnetic field, is sufficient for full applicability
of the available stochastic theory of radiation for the case under
study. This condition means that the distortion of the electron
trajectory due to the regular magnetic field is small during  the
period of electric oscillations, thus, it is not needed to be small
at any spatial scale, so no further restriction on values of
$\omega_0$, $\omega\st$, and $\omega\pe$ is necessary.

\begin{figure}
\hspace{-0.4in}
\includegraphics[bb=18 5 282 217,height=2.5in]{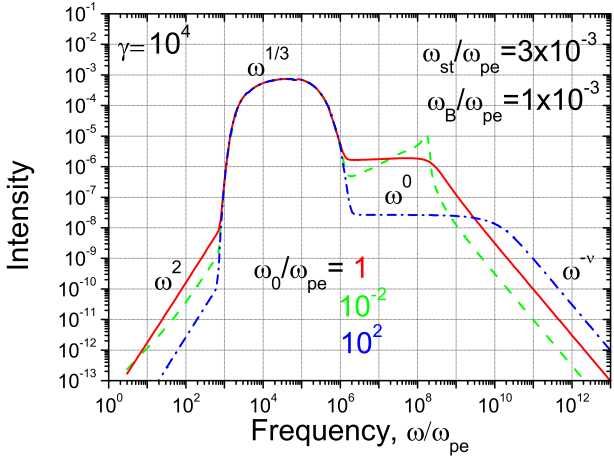} 
\qquad
\includegraphics[bb=18 5 282 217,height=2.5in]{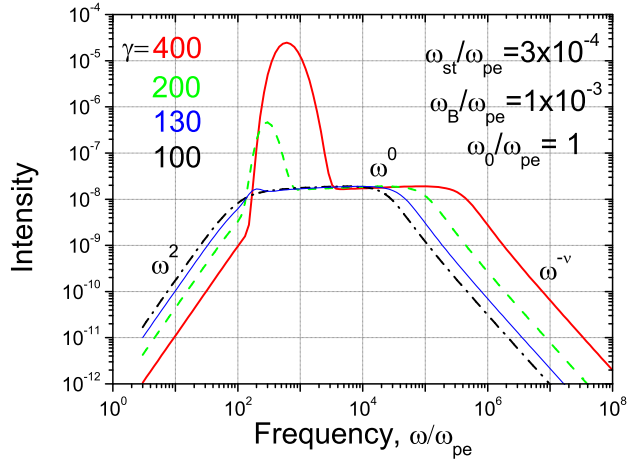}
\caption{\small \it DRL spectra in the presence of uniform magnetic
field with the energy density of the order of 10$\%$ of the Langmuir
turbulence energy density. DRL dominates at frequencies above
$\omega_{Be}\gamma^2$, where the perturbative treatment is
applicable as long as $\omega\st \ll \omega\pe$. }\label{lang_large}
\caption{\small \it DRL spectra in the presence of uniform magnetic
field.  The Langmuir turbulence energy density is of the order of
10$\%$ of the magnetic energy density. Synchrorton spectrum is
strongly suppressed by the effect of density. DRL dominates the
entire spectrum for sufficiently low-energy electrons with
$\gamma<120$.  } \label{DRL_Razin}
\end{figure}

Figure \ref{lang_large} shows examples  of the spectra for the cases
of short-wave or long-wave Langmuir turbulence superimposed on a
regular magnetic field. The full spectrum consists of the standard
synchrotron contribution (region $I_\omega \propto \omega^{1/3}$
with exponential cut-offs at low- and high-frequency edges) and DRL
contribution, which is the most prominent at the high frequencies
$\omega \gg \omega\Be\gamma^2$, although it is also present at
sufficiently low frequencies, where the spectrum $I_\omega \propto
\omega^2$ is formed.

Figure \ref{lang_large} allows direct comparison of the DRL spectrum
with that of the standard synchrotron radiation. We point out that
these two emission processes occupy distinct frequency ranges if
$\omega\Be \ll \omega\pe$. In particular, the synchrotron spectrum
displays exponential cut-off at the frequencies $\omega > \omega\Be
\gamma^2$, while the DRL displays here flat of even rising spectrum
up to $\omega \sim \omega\pe \gamma^2$. This means that DRL can
dominate this spectral range even if the energy density in the
Langmuir turbulence is lower than the magnetic field energy density.
This holds also for a power-law energy spectrum with a high-energy
cut-off at some $\gamma\mmax$ for the spectral range $\omega >
\omega\Be \gamma\mmax^2$.

Moreover, in sufficiently dense plasmas DRL can dominate the entire
radiation spectrum even for the conditions when the Langmuir
turbulence energy density is much smaller the the magnetic energy
density. This happens for relatively low-energy (although
ultrarelativistic) electrons, whose synchrotron emission is
significantly suppressed by the Razin-effect (referred also to as
density effect). Figure \ref{DRL_Razin} displays such an example.
Here the turbulent energy density is ten times lower than the
magnetic energy density. Radiation spectrum produced by higher
energy electrons ($\gamma>120$ for the parameters selected to plot
the figure) consists of both DRL and synchrotron contribution,
although the later becomes narrower and weaker as $\gamma$
decreases. Eventually, for $\gamma<120$, the synchrotron
contribution disappears being strongly suppressed by the Razin
effect in contrast to the DRL spectrum, which is less sensitive to
the density effect.

\section{Discussion}


Shock waves, in particular, jet shocks are believed to be the
sites where various kinds of the two-stream instability can be
operational. Depending on the conditions at the shock front and
its vicinity either magnetic or electric fluctuations or both will
be excited. Here, we specifically considered the case when random
electric fields in a form of Langmuir wave turbulence dominate.

Modern computer simulations of shock wave interactions, especially
in the relativistic case, suggest that the energy density of the
excited Langmuir turbulence can be far in excess of the energy of
the initial regular magnetic field. In particular, at the wave front
the electric field can be as strong as the corresponding
wave-breaking limit, i.e., $\omega\st \sim \omega\pe$. In this case
the random walk of relativistic electrons in the stochastic electric
field can give rise to powerful contribution in the nonthermal
emission of an astrophysical object, entirely dominating full
radiation spectrum or some broad part of it.

So far, the DRL (electrostatic bremsstrahlung) has been applied to a
number of astrophysical objects. For example,
\cite{Schlickeiser_2003} noted that electrostatic bremsstrahlung is
an attractive alternative to standard synchrotron radiation to
produce the observed nonthermal emission from jets in active
galactic nuclei. In addition, \cite{Schroeder_etal_2005} developed a
simplified model of the galactic diffuse sub-MeV emission based on
monochromatic approximation of both synchrotron radiation and the
DRL, which gives rise to a remarkably good agreement between the
model and the observations. We point out that the use of presented
here DRL spectra will be helpful to further develop that model
especially in the range of high-energy cut-off of the radiation
spectra.

Although any detailed application of the considered emission process
is beyond the scope of this paper, we mention that the DRL is also
a promising mechanism for the gamma-ray bursts and extragalactic
jets. In particular, some of the prompt gamma-ray bursts display
rather hard low-energy spectra with the \emph{photon} spectral index
$\alpha$ up to 0. The DRL spectral asymptote $I_\omega \propto
\omega^{1}$, which appears just below the spectral peak at
$2\omega\pe\gamma^2$, fits well to those spectra. Remarkably, the
flat lower-frequency asymptote, $I_\omega \propto \omega^{0}$, can
account for the phenomenon of the X-ray excess
\citep{Preece_etal_1996,Sakamoto_etal_2005} and prompt optical
flashes accompanying some GRBs.

In addition, this mechanism \citep[along with the DSR in random
magnetic fields,][]{Fl_2005c} can be relevant to the UV-X-ray
flattenings observed in some extragalactic jets. For example,
although full spatially resolved radio to X-ray spectra of the jet
in M87 agrees well with the DSR model \citep{Fl_2005c}, for the jet
in 3C~273 this agreement holds from the radio to UV range, while its
X-ray emission seems to require an additional component
\citep{Jester_etal_2006}. Alternatively, the entire UV-to-X-ray
spectrum of 3C~273 might be produced by DRL, which can be much
flatter than usual DSR  (see, e.g., Figure \ref{lang_large}) in the
range $\omega\Be\gamma^2 \ll \omega \ll \omega\pe\gamma^2$.

\section*{Acknowledgments}

This work was supported in part by the RFBR grants 06-02-16295 and
06-02-16859. We have made use of NASA's Astrophysics Data System
Abstract Service. We are grateful to the reviewer, Dr. R.
Schlickeiser, for his valuable comments to the paper.


\bibliographystyle{mn2e} \bibliography{DSR_PWNs,DSR_Langmuir}

\label{lastpage}

\end{document}